\documentclass[11pt,letterpaper]{article}

\usepackage{amsmath,amssymb}
\usepackage{xcolor}

\newtheorem{lemma}{Lemma}
\newtheorem{theorem}{Theorem}
\newtheorem{problem}{Problem}
\newtheorem{corollary}{Corollary}
\newtheorem{definition}{Definition}

\newtheorem{observation}{Observation}

\title{Beyond the Longest Letter-duplicated Subsequence Problem}
\author{
Wenfeng Lai \footnote{College of Computer Science and Technology, Shandong University, Qingdao, China.
Email: \texttt{2290892069@qq.com}.}
\and Adiesha Liyanage \footnote{Gianforte School of Computing, Montana State University, Bozeman, MT 59717, USA.
Email: \texttt{adiesha@gmail.com}.}
\and Binhai Zhu \footnote{Corresponding author; Gianforte School of Computing, Montana State University, Bozeman, MT 59717, USA.
Email: \texttt{bhz@montana.edu}.}
\and Peng Zou \footnote{Gianforte School of Computing, Montana State University, Bozeman, MT 59717, USA.
Email:\texttt{peng.zou@student.montana.edu}.}
}

\date{}

\begin{document}

\maketitle

\begin{abstract}
Motivated by computing duplication patterns in sequences, 
a new fundamental problem called the longest letter-duplicated subsequence (LLDS) is proposed.
Given a sequence $S$ of length $n$, a letter-duplicated subsequence is a subsequence of $S$ in the form of $x_1^{d_1}x_2^{d_2}\cdots x_k^{d_k}$ with $x_i\in\Sigma$, $x_j\neq x_{j+1}$ and $d_i\geq 2$ for all $i$ in $[k]$ and $j$ in $[k-1]$. 
A linear time algorithm for computing the longest letter-duplicated subsequence (LLDS) of $S$ can be easily obtained. In this paper, we focus on two variants of this problem. We first consider the constrained version when $\Sigma$ is unbounded, each letter appears in $S$ at least 6 times and all the letters in $\Sigma$ must appear in the solution. We show that the problem is NP-hard (a further twist indicates that the problem does not admit any
polynomial time approximation). The reduction is from possibly the simplest version of SAT that is NP-complete, $(\leq 2,1,\leq 3)$-SAT, where each variable appears at most twice positively and exact once negatively, and each clause contains at most three literals and some clauses must contain exactly two literals. (We hope that this technique will serve as a general tool to help us proving the NP-hardness for some more tricky sequence problems involving only one sequence --- much harder than with at least two input sequences, which we apply successfully at the end of the paper on some extra variations of the LLDS problem.)  We then show that when each letter appears in $S$ at most 3 times, then the problem admits a factor $1.5-O(\frac{1}{n})$ approximation. Finally, we consider the weighted version, where the weight of a block $x_i^{d_i} (d_i\geq 2)$ could be any positive function which might not grow with $d_i$. We give a non-trivial $O(n^2)$ time dynamic programming algorithm for this version, i.e., computing an LD-subsequence of $S$ whose weight is maximized.\\

\end{abstract}

\section{Introduction}

In biology, duplication is an important part of evolution. There are two kinds of duplications: arbitrary segmental duplications (i.e., select a segment and paste it somewhere else) and tandem duplications (which is in the form of $X\rightarrow XX$, where $X$ is any segment of the input sequence).
It is known that the former duplications occur frequently in cancer genomes \cite{Sharp05,CGARN11,Ciriello13}. On the other hand, the latter are common under different scenarios, for example, it is known that the tandem duplication of
3 nucleotides {\tt CAG} is closely related to the Huntington disease \cite{Macdonald93}. In addition, tandem duplications can occur at the genome level (acrossing different genes) for certain types of cancer
\cite{Oesper12}. In fact, as early as in 1980, Szostak and Wu provided evidence that gene duplication is the main driving force behind evolution, and the majority of duplications are tandem \cite{uneqcrnature}. Consequently, it was not a surprise that in the first sequenced human genome around 3\% of the genetic contents are in the form of tandem repeats \cite{Lander01}.

Independently, tandem duplications were also studied in {\em copying systems} \cite{Ehren84}; as well as in formal languages \cite{Bovet92,Dassow99,Wang00}. In 2004, Leupold et al. posed a fundamental question regarding tandem duplications: what is the complexity to compute the minimum tandem duplication distance between two sequences $A$ and $B$ (i.e., the minimum number of tandem duplications to convert $A$ to $B$). In 2020, Lafond et al. answered this open question by proving that this problem is NP-hard for an unbounded alphabet \cite{DBLP:conf/stacs/LafondZZ20}. In fact, Lafond et al. proved in the journal version that the problem is NP-hard even if $|\Sigma|\geq 4$ by encoding each letter in the unbounded alphabet proof with a square-free string over a new alphabet of size 4 (modified from Leech's construction \cite{leech57}), which covers the case most relevant with biology, i.e., when $\Sigma=\{{\tt A},{\tt C},{\tt G},{\tt T}\}$ \cite{Lafond21}. Independently, Cicalese and Pilati showed that the problem is NP-hard for $|\Sigma|=5$ using a different encoding method \cite{DBLP:conf/iwoca/CicaleseP21}.

Motivated by the above applications (especially when some mutations occur after the duplications), some new problems related to duplications are proposed and studied in this paper. Given a sequence $S$ of length $n$,
a letter-duplicated subsequence (LDS) of $S$ is a subsequence of $S$ in the form $x_1^{d_1}x_2^{d_2}\cdots x_k^{d_k}$ with $x_i\in\Sigma$, where $x_j\neq x_{j+1}$ and $d_i\geq 2$ for all $i$ in $[k]$ and $j$ in $[k-1]$. (Each $x_i^{d_i}$ is called an LD-block.) Naturally, the problem of computing the longest letter-duplicated subsequence (LLDS) of $S$ can be defined, and a simple linear time algorithm can be obtained. 
(We comment that recently a similar problem called {\em longest run subsequence} was studied \cite{DBLP:conf/wabi/SchrinnerGWSSK20}, it differs from our problem in that each letter appears consecutively at most once in the solution and does not have to be repeated, and the goal is the same, i.e., the length of the subsequece
is to be maximized.)

In this paper, we focus on some important variants around the fundamental LLDS problem, focusing on the constrained and weighted cases. The constraint is to demand that all letters in $\Sigma$ appear in a resulting LDS, which simulates that in a genome with duplicated genes, how to compute the maximum duplicated pattern while including all the genes. Then we have two problems: feasibility testing (FT for short, which decides whether an
LDS of $S$ containing all letters in $\Sigma$ exists) and the problem of maximizing the length of a resulting LDS where all letters in the alphabet appear, which we call LLDS+. It turns out that the status of these two problems change quite a bit when $d$, the maximum number a letter can appear in $S$, varies. We denote the corresponding problems as $FT(d)$ and \emph{LLDS+}$(d)$ respectively. Let $|S|=n$, we summarize our main results in this paper as follows:
\begin{enumerate}
    \item We show that when $d\geq 6$, both $FT(d)$ and (the decision version of) \emph{LLDS+}$(d)$ are NP-complete, which implies that \emph{LLDS+}$(d)$ does not have an polynomial-time approximation algorithm when $d\geq 6$.
    \item We show that when $d=3$, $FT(d)$ is decidable in $O(n^2)$ time, which implies that \emph{LLDS+}$(3)$ admits a factor-1.5 approximation. With an increasing running time, we could improve the factor to $1.5-O(\frac{1}{n})$.
    \item When a weight of an LD-block is any positive function (i.e., it does not even have to grow with its length), we present a non-trivial $O(n^2)$ time dynamic programming problem for this Weighted-LDS problem.
\end{enumerate}

In the literature, the only known related work is to compute the longest
{\em square} subsequence of an input sequence $S$, for which Kosowski gave an $O(n^2)$ time algorithm \cite{DBLP:conf/spire/Kosowski04}. At the end of
paper, we will briefly mention two extra variations of the LLDS problem,
where in the solution, i.e., a subsequence of $S$ in the form of $x_1^{d_1}x_2^{d_2}\cdots x_k^{d_k}$, each $x_i$ is either a substring or a subsequence of $S$.  Then, what Kosowski considered is the more restricted
version of the latter, i.e., $x_1^{d_1}x_2^{d_1}$, with $x_1=x_2$ and $d_1=d_2$.

This paper is organized as follows. In section 2 we give necessary definitions. In section 3 we focus on showing that the LLDS+ and FT problems are NP-complete when $d\geq 6$ and some positive results when $d=3$. In section 4 we give polynomial-time algorithms for
Weighted-LDS. We conclude the paper in section 5.

\section{Preliminaries}

Let $\mathbb{N}$ be the set of natural numbers. For $q\in\mathbb{N}$, we use $[q]$ to represent the set $\{1,2,...,q\}$.
Throughout this paper, a sequence $S$ is over a finite alphabet $\Sigma$.
We use $S[i]$ to denote the $i$-th letter in $S$ and $S[i..j]$ to denote
the substring of $S$ starting and ending with indices $i$ and $j$ respectively. (Sometimes we also use $(S[i],S[j])$ as an interval representing the substring $S[i..j]$.) With the standard run-length representation, $S$ can be represented as
$y_1^{a_1}y_2^{a_2}\cdots y_q^{a_q}$, with $y_i\in\Sigma,y_j\neq y_{j+1}$ and $a_j\geq 1$, for $i\in[q],j\in[q-1]$. If a letter $a$ appears multiple times
in $S$, we could use $a^{(i)}$ to denote the $i$-th copy of it (reading from left to right). Finally, a {\em subsequence} of $S$ is a string obtained by deleting some letters in $S$.

\subsection{The LLDS Problem}

A subsequence $S'$ of $S$ is a letter-duplicated subsequence (LDS) of $S$ if it is in
the form of $x_1^{d_1}x_2^{d_2}\cdots x_k^{d_k}$, with $x_i\in\Sigma,x_j\neq x_{j+1}$ and $d_i\geq 2$, for $i\in[k],j\in[k-1]$. We call each $x_i^{d_i}$ in $S'$ a {\em letter-duplicated block} (LD-block, for short).
For instance, let $S=abcacabcb$, then $S_1=aaabb$, $S_2=ccbb$ and $S_3=ccc$ are all letter-duplicated subsequences of $S$, where $aaa$ and $bb$ in $S_1$, $cc$ and $bb$ in $S_2$, and $ccc$ in $S_3$ all form the corresponding LD-blocks. Certainly, we are interested in the longest ones --- which gives us the longest letter-duplicated subsequence (LLDS) problem. 

As a warm-up, we solve this
problem by dynamic programming. We first have the following observation.

\begin{observation}
Suppose that there is an optimal LLDS solution for a given sequence $S$ of length $n$, in the form of $x_1^{d_1} x_2^{d_2} \ldots x_k^{d_k}$. Then it is
possible to decompose it into a generalized LD-subsequence $y_1^{e_1} y_2^{e_2} \ldots y_p^{e_p}$, where 
\begin{itemize}
    \item $ 2 \leq e_i \leq 3$, for $i\in[p]$,
    \item $p \geq k$,
    \item $y_j$ does not have to be different from $y_{j+1}$, for $j\in[p-1]$.
\end{itemize}
\label{prop:decomp}
\end{observation}

The proof is straightforward:
For any natural number $\ell \geq 3$, we can decompose it as $\ell = \ell_1 + \ell_2 + \ldots + \ell_z \geq 3$, such that $2 \leq \ell_j \leq 3$ for $1 \leq j \leq z$. Consequently, for every $d_i>3$, we could decompose it into a sum
of 2's and 3's. Then, clearly, given a generalized LD-subsequence, we could easily obtain the corresponding LD-subsequence by combining $y_i^{e_i}y_{i+1}^{e_{i+1}}$ when
$y_i=y_{i+1}$.

We now design a dynamic programming algorithm for LLDS.
Let $L(i)$ be the length of the optimal LLDS solution for $S[1..i]$. The recurrence for $L(i)$ is as follows.
\begin{align*}
L(0) & = 0, \\
L(1) & = 0, \\
L(i) & = \max 
\begin{cases}
L(i-x-1) + 2 &  x = \min \{x| S[i-x] = S[i] \}, x\in(0,i-1]  \\
L(i-x) + 1 & x = \min \{x| S[i-x] = S[i] \}, x\in (0,i-1] \\
L(i-1) & \text{otherwise.}
\end{cases}
\end{align*}

Note that the step involving $L(i-x)+1$ is essentially a way to handle a generalized LD-subsequence of length 3 (by keeping $S[i-x]$ for the next level computation) and cannot be omitted following the above observation. For instance, if $S=dabcdd$ then without that step we would miss the
optimal solution $ddd$.

The value of the optimal LLDS solution for $S$ can be found in $L(n)$. For the running time, for each $S[x]$ we just need to scan $S$ to find the closest $S[i]$ such that
$S[x]=S[i]$. With this information, the table $L$ can be filled in linear time. With a simple augmentation, the actual sequence corresponding to $L(n)$ can also be found in linear time. Hence LLDS can be solved in $O(n)$ time.

\subsection{The Variants of LLDS}

In this paper, we focus on the following variations of the LLDS problem.

\begin{definition}
\textbf{\emph{Constrained Longest Letter-Duplicated Subsequence}}\\ ($LLDS+$ for short)

{\bf Input}: A sequence $S$ with length $n$ over an alphabet $\Sigma$ and an integer $\ell$.

{\bf Question}: Does $S$ contain a letter-duplicated subsequence $S'$ with length at least $\ell$ such that all letters in $\Sigma$ appear in $S'$?
\end{definition}

\begin{definition}
\textbf{\emph{Feasibility Testing}} (FT for short)

{\bf Input}: A sequence $S$ with length $n$ over an alphabet $\Sigma$.

{\bf Question}: Does $S$ contain a letter-duplicated subsequence $S''$ such that all letters in $\Sigma$ appear in $S''$?
\end{definition}

For LLDS+ we are really interested in the optimization version, i.e., to maximize $\ell$. Note that, though looking similar, FT and the decision version of LLDS+ are different: if there is no feasible solution for
FT, certainly there is no solution for LLDS+; but even if there is a feasible
solution for FT, computing an optimal solution for LLDS+ could still be non-trivial.

Finally, let $d$ be the maximum number of times a letter in $\Sigma$ appears in $S$. Then, we can represent the corresponding versions for LLDS+ and FT as
\emph{LLDS+}$(d)$ and $FT(d)$ respectively.

It turns out that (the decision version of) \emph{LLDS+}$(d)$ and $FT(d)$ are both NP-complete when $d\geq 6$, while when $d=3$ the status varies: $FT(3)$ can be decided in $O(n^2)$ time, which immediately implies that \emph{LLDS+}$(3)$ has a factor-1.5 approximation. (If we are willing to increase the running time --- still polynomial but higher than $O(n^2)$, with some simple twist we could improve the approximation factor for \emph{LLDS+}$(3)$ to $1.5-O(\frac{1}{n})$.)
We present the details in the next section. In Section 4, we will consider
an extra version of LLDS, Weighted-LDS, where the weight of an LD-block
is an arbitrary positive function.

\section{Hardness with the full-appearance constraint}

\subsection{Hardness for LLDS+($d$) and FT($d$) when $d\geq 6$}
We first try to prove the NP-completeness of the (decision version of) LLDS+($d$), when $d\geq 6$.
In fact, we need a very special version of SAT, which is possibly the simplest version of SAT remaining NP-complete.

Given a 3SAT formula $\phi$, which is a conjunction of $m$ disjunctive clauses (over $n$ variable $x_i$'s), each clause $F_j$ containing exactly 3 literals (i.e., in the form of $x_i$ or $\bar{x}_i$), the problem is to find whether there is a satisfiable truth assignment for $\phi$.

\begin{definition}
$(\leq 2,1,\leq 3)$-SAT: this is a special case of SAT where each variable $x_i$ appears at most twice and $\bar{x}_i$ appears exactly once in $\phi$; moreover, each clause
contains either two or three literals (which will be called 2-clause and 3-clause henceforth).
\end{definition}
\begin{lemma}
$(\leq 2,1,\leq 3)$-SAT is NP-complete.
\end{lemma}

{\bf Proof.} We modify the proof by Tovey \cite{DBLP:journals/dam/Tovey84}. Given a 3SAT formula $\phi$, without loss of generality, assume that each variable $x_i$ and its complement $\bar{x}_i$ appears in (different clauses of) $\phi$. We convert $\phi$ to $\phi'$ in the form of $(\leq 2,1\leq 3)$-SAT as follows.
\begin{itemize}
    \item if both $x_i$ and $\bar{x}_i$ appears once in $\phi$, do nothing.
    \item if $x_i$ appears twice and $\bar{x}_i$ appears once in $\phi$, do nothing.
    \item if $\bar{x}_i$ appears twice and $x_i$ appears once in $\phi$, replace $\bar{x}_i$ with a new variable $z$ and replace $x_i$ by $\bar{z}$.
    \item Otherwise, if the total number of literals of $x_i$ (i.e., $x_i$ and $\bar{x}_i$) is $k\geq 4$ then introduce $k$ variables $y_{i,1},y_{i,2},\cdots,y_{i,k}$ replacing the $k$ literals of $x_i$ respectively. Moreover, let $z_{i,j}$ be $y_{i,j}$ if the $j$-th literal of $x_i$ is $x_i$ and let $z_{i,j}$ be $\bar{y}_{i,j}$ if the $j$-th literal of $x_i$ is $\bar{x}_i$. Finally, add $k$ 2-clauses as
    $(z_{i,j}\vee \bar{z}_{i,j+1})$ for $j=1..k-1$ and $(z_{i,k}\vee \bar{z}_{i,1})$. (Note that it always holds that $\overline{\bar{z}}=z$.)
\end{itemize}
Following \cite{DBLP:journals/dam/Tovey84}, when $k\geq 4$, the 2-clauses
added will force all $z_{i,j}$'s to have all {\tt True} values or all {\tt False} values. (The only difference between our construction and Tovey's is that all literals appearing at least 4 times in the original clauses in $\phi$ are replaced by positive
variables in the form of $y_{i,j}$'s; the negated literal $\bar{y}_{i,j}$ could only occur in the newly created 2-clauses --- exactly once for each $y_{i,j}$. On the other hand, each $y_{i,j}$ occur twice --- once in the original 3-clauses of $\phi$ and once in the newly created 2-clauses.)
It is obvious to see that $\phi$ is satisfiable if and only if $\phi'$ is
satisfiable. The transformation obviously takes $O(|\phi|)$ time. Hence the lemma is proven.
\hfill $\Box$

We comment that $(\leq 2,1\leq 3)$-SAT, while seemingly similar to SAT3W (each clause has at most 3 literals and each clause has at most one negated variable \cite{DBLP:conf/stoc/Schaefer78}), is in fact different from it. (Following the Dichotomy Theorem for SAT by Schaefer \cite{DBLP:conf/stoc/Schaefer78}, SAT3W is in P.) The difference is that in $\phi'$ we could even have a clause containing 3 negated
variables.

Now let $\phi$ be an instance of $(\leq 2,1,\leq 3)$-SAT where either $x_i$ and $\bar{x}_i$ appears once in $\phi$ (we call such an $x_i$ a {\em (1,1)-variable}), or $x_i$ appears twice and $\bar{x}_i$ appears once in $\phi$ (we call such an $x_i$ a {\em (2,1)-variable}), for $i=1..n$.
Without loss of generality, we assume $\phi=F_1\wedge F_2\wedge \cdots \wedge F_m$ and there are $n$ variables $x_1,x_2,\cdots,x_n$; moreover, we assume that $F_j$ cannot contain $x_i$ and $\bar{x}_i$ at the same time.
Given $F_j$ we say $F_jF_j$ forms a {\em 2-duplicated clause-string}.

Given a {\em (2,1)-sequence} $T=ACABCB$ over $\{A,B,C\}$, where $A,B$ and $C$ all appear twice, it is easy to verify that the maximal LD-subsequences of $T$ are $AABB$ or $CC$.
Similarly, given a {\em (1,1)-sequence} $T=ACCA$ over $\{A,C\}$, where $A$ and $C$ both appear twice, it is easy to see that the maximal (longest) LD-subsequences of $T$ are $AA$ or $CC$.

For each (1,1)-variable $x_i$, i.e., $x_i$ and $\bar{x}_i$ appears once in $\phi$, say $x_i$ in $F_j$ and $\bar{x}_i$ in $F_k$, we define $L_i$ as a (1,1)-sequence: $F_jF_kF_kF_j$.
For each (2,1)-variable $x_i$, i.e., $x_i$ appears twice and $\bar{x}_i$ appears once in $\phi$, say $x_i$ in $F_j$ and $F_k$, and $\bar{x}_i$ in $F_\ell$, we define $L_i$ as a (2,1)-sequence: $F_jF_{\ell}F_jF_kF_{\ell}F_k$. 

Now we proceed to construct the sequence $S$ from an $(\leq 2,1,\leq 3)$-SAT instance $\phi$.

$$S=g_1g_1L_1g_2g_2\cdots g_ig_iL_i\cdots g_{n-1}g_{n-1}L_{n-1}g_ng_nL_ng_{n+1}g_{n+1}.$$

We claim the following: $\phi$ is satisfiable if and only if LLDS+ has a solution of length at least $2(n+1)+4K_1+2K_2+2J$, where $K_1,K_2$ are the number of (2,1)-variables in $\phi$ which are assigned {\tt True} and {\tt False} respectively
and $J$ is the number of (1,1)-variables in $\phi$.

{\bf Proof.} ``Only-if part": Suppose that $\phi$ is satisfiable. If a (1,1)-variable $x_i$ is assigned {\tt True}, to have a solution for LLDS+, in $L_i$ we select the 2-duplicated clause-string $F_jF_j$; likewise, if $x_i$ is assigned {\tt False} we select $F_kF_k$ instead. Similarly, if a (2,1)-variable $x_i$ is assigned {\tt True}, to have a solution for LLDS+, in $L_i$ we select two 2-duplicated clause-strings $F_jF_jF_kF_k$; likewise, if $x_i$ is assigned {\tt False} we select $F_{\ell}F_{\ell}$.
Since $g_ig_i$ only occurs once in $S$ and $T$, we must include them in the solution. Clearly we have a solution for LLDS+ with length $2(n+1)+4K_1+2K_2+2J$.

``If part": If LLDS+ has a solution of length at least $2(n+1)+4K_1+2K_2+2J$, by definition,
it must contain all $g_ig_i$'s. To find the truth assignment, we look at the contents between $g_{i}g_{i}$ and $g_{i+1}g_{i+1}$ in the solution as well as in $S$ (i.e., $L_i$). If $x_i$ is a (1,1)-variable, $L_i=F_jF_kF_kF_j$ and in the solution $F_jF_j$ is kept then we assign $x_i\leftarrow {\tt True}$; otherwise, we assign $x_i\leftarrow {\tt False}$. If $x_i$ is a (2,1)-variable, $L_i=F_jF_{\ell}F_jF_kF_{\ell}F_k$ and in the solution either $F_jF_jF_kF_k$, $F_jF_j$ or $F_kF_k$ is kept then we assign $x_i\leftarrow {\tt True}$. (When $F_jF_j$ or $F_kF_k$ is kept, then the LLDS+ solution could be longer by augmenting this sub-solution to $F_jF_jF_kF_k$.) If in the solution $F_{\ell}F_{\ell}$ is kept instead then we assign $x_i\leftarrow {\tt False}$. Since all clauses must appear in a solution of LLDS+, clearly $\phi$ is satisfied.
\hfill $\Box$

We comment that $2(n+1)+4K_1+2K_2+2J=2(n+1)+2K_1+2n=4n+2+2K_1$, as $K_1+K_2+J=n$. (Note that $K_1$ only
represents a part of the truth assignment for $\phi$ and it could be general, i.e., $K_1$ could be $\Omega(n)$.) But the former
makes our arguments more clear. This reduction obviously takes $O(m+n)$ time. Note that each 3-clause $F_j$ appears 6 times in $S$ and each 2-clause $F_\ell$ appears 4 times in $S$ respectively, while each $g_k, k\in [n+1]$, appears twice in $S$. Since we could arbitrarily add an LD-block $u^j$, with $u\not\in\Sigma$ and $j\geq 6$, at the end
of $S$, we have the following theorem.

\begin{theorem}
The decision version of LLDS+($d$) is NP-complete for $d\geq 6$.
\end{theorem}

We next present an example for this proof.

Example. Let $\phi=F_1\wedge F_2\wedge F_3\wedge F_4\wedge F_5$ $=(x_1\vee x_2\vee x_3)\wedge (\bar{x}_1\vee \bar{x}_2\vee x_3)\wedge (x_2\vee \bar{x}_3\vee x_4)\wedge (x_1\vee x_4\vee x_5)\wedge (\bar{x}_4\vee \bar{x}_5)$. Then
$$S=g_1g_1F_1F_2F_1F_4F_2F_4\cdot g_2g_2 F_1F_2F_1F_3F_2F_3\cdot g_3g_3 F_1F_3F_1F_2F_3F_2$$
$$\cdot g_4g_4F_3F_5F_3F_4F_5F_4\cdot g_5g_5 F_4F_5F_5F_4\cdot g_6g_6,$$
Corresponding to the truth assignment, $x_1,x_4={\tt True}$ and $x_2,x_3,x_5={\tt False}$, we have
$$S'=g_1g_1F_1F_1F_4F_4\cdot g_2g_2F_2F_2\cdot g_3g_3F_3F_3\cdot g_4g_4F_3F_3F_4F_4\cdot g_5g_5F_5F_5\cdot g_6g_6,$$
which is of length $2(5+1)+4\times K_1+2\times K_2+2\times 1=12+4\times 2 + 2\times 2+ 2=26$.

The above theorem implies the following corollary.
\begin{corollary}
FT($d$) is NP-complete for $d\geq 6$.
\end{corollary}

{\bf Proof.} The reduction remains the same. We just need to augment the proof in the reverse direction.
Suppose there is a feasible solution $S''$ for $S$ for the feasibility testing problem.
Again, all $g_ig_i$'s must be in $S''$. We now look at the contents between $g_{i}g_{i}$ and $g_{i+1}g_{i+1}$ in $S$ (i.e., $L_i$) and $S''$. Corresponding to $L_i$, if in $S''$ we have an empty string between $g_{i}g_{i}$ and $g_{i+1}g_{i+1}$, then we can assign $x_i$ either as {\tt True} or {\tt False}.
If $L_i=F_jF_kF_kF_j$, i.e., $x_i$ is a (1,1)-variable, 
and $F_jF_j$ is kept in $S''$ then we assign $x_i\leftarrow {\tt True}$; otherwise, we assign $x_i\leftarrow {\tt False}$. If 
$L_i=F_jF_{\ell}F_jF_kF_{\ell}F_k$, i.e., $x_i$ is a (2,1)-variable, 
and either $F_jF_jF_kF_k$, $F_jF_j$ or $F_kF_k$ is kept in $S''$ then we assign $x_i\leftarrow {\tt True}$. If in the solution $F_{\ell}F_{\ell}$ is kept instead then we assign $x_i\leftarrow {\tt False}$. By definition, all clauses must appear in $S''$ (solution of FT), clearly $\phi$ is satisfied.
It is clear that FT belongs to NP as a solution can be easily checked in polynomial time.
\hfill $\Box$

The above corollary essentially implies that the optimization version of \emph{LLDS+}$(d)$, $d\geq 6$, does not
admit any polynomial-time approximation algorithm (regardless of the approximation factor),
since any such approximation would have to return
a feasible solution. A natural direction to approach LLDS+ is to design a bicriteria approximation for LLDS+, where
a factor-$(\alpha,\beta)$ bicriteria approximation algorithm is a polynomial-time algorithm which
returns a solution of length at least ${OPT}/\alpha$ and includes at least $N/\beta$ letters,
where $N=|\Sigma|$ and {OPT} is the optimal solution value of LLDS+. We show that obtaining a 
bicriteria approximation algorithm for LLDS+ is no easier than approximating LLDS+ itself. 

\begin{corollary}
If \emph{LLDS+}$(d), d\geq 6$, admitted a factor-$(\alpha,N^{1-\epsilon})$ bicriteria approximation for any $\epsilon<1$, then \emph{LLDS+}$(d), d\geq 6$, would also admit a factor-$\alpha$ approximation, where $N$ is the alphabet size.
\end{corollary}

{\bf Proof.} Suppose that a factor-$(\alpha,N^{1-\epsilon})$ bicriteria approximation algorithm ${\cal A}$ exists.
We construct an instance $S^*$ for \emph{LLDS+}(6) as follows. (Recall that $S$ is the sequence we constructed
from an $(\leq 2,1\leq 3)$-SAT instance $\phi$ in the proof of Theorem 1.)
In addition to $\{F_i|i=1..m\}\cup \{g_j|j=1..n+1\}$ in the alphabet,
we use a set of integers $\{1,2,...,(m+n+1)^x-(m+n+1)\}$, where $x$ is some integer to be determined.
Hence, $$\Sigma=\{F_i|i=1..m\}\cup \{g_j|j=1..n+1\}\cup \{1,2,...,(m+n+1)^x-(m+n+1)\}.$$
We now construct $S^*$ as
$$S^*=1\cdot 2\cdots ((m+n+1)^x-(m+n+1))\cdot S\cdot ((m+n+1)^x-(m+n+1))$$
$$\cdot ((m+n+1)^x-(m+n+1)-1)\cdots 2\cdot 1.$$
Clearly, any bicriteria approximation for $S^*$ would return an approximate solution for $S$ as including any number in $\{1,2,...,(m+n+1)^x-(m+n+1)\}$ would result in a solution of size only  2.

Notice that we have $N=m+(n+1)+(m+n+1)^x-(m+n+1)=(m+n+1)^x$. In this case, the fraction of letters in $\Sigma$ that is used to form
such an approximate solution satisfies
$$\frac{m+(n+1)}{(m+n+1)^x}\leq \frac{1}{N^{1-\epsilon}},$$
which means it suffices to choose $x\geq \lceil 2-\epsilon\rceil=2$.
\hfill $\Box$

\subsection{Solving the Feasiblility Testing Version for $d=3$}

For the Feasibility Testing Version, as mentioned earlier, Corollary 1 implies that the problem
is NP-complete when $d\geq 6$. We next show that if $d=3$, then the problem can be decided in
polynomial time.
\begin{lemma}
Given a string $S$ over $\Sigma$ such that each letter in $S$ appears at most 3 times, if a feasible
solution for $FT(3)$ contains a 3-block then there is a feasible solution for $FT(3)$ which only
uses 2-blocks.
\end{lemma}

{\bf Proof.} Suppose that $S=\cdots a^{(1)}\cdots a^{(2)}\cdots a^{(3)}\cdots$, and $a^{(1)}a^{(2)}a^{(3)}$ is
a 3-block in a feasible solution for $FT(3)$. (Recall that the superscript only indicates the appearance order of letter $a$.) Then we could replace $a^{(1)}a^{(2)}a^{(3)}$
by either $a^{(1)}a^{(2)}$ or $a^{(2)}a^{(3)}$. The resulting solution is still a feasible
solution for $FT(3)$.
\hfill $\Box$

Lemma 2 implies that the $FT(3)$ problem can be solved using 2-SAT. For each letter $a$, we denote the interval $(a^{(1)},a^{(2)})$ as a variable $v_a$, and we denote $(a^{(2)},a^{(3)})$ as $\bar{v}_a$. (Clearly one cannot select $a^{(1)}a^{(2)}$ and $a^{(2)}a^{(3)}$ as 2-blocks at the same time.) Then,
if another interval $(b^{(1)},b^{(2)})$ overlaps the interval $(a^{(1)},a^{(2)})$, we have a 2-SAT
clause $\overline{v_a\wedge v_b}=(\bar{v}_a\vee\bar{v}_b)$. Forming a 2-SAT instance $\phi"$ for
all such overlapping intervals and it is clear that we can decide whether $\phi"$ is satisfiable in $O(n^2)$ time (as we could have $O(n^2)$ pairs of overlapping intervals).

\begin{theorem}
Let $S$ be a string of length $n$. Whether $FT(3)$ has a solution can be decided in $O(n^2)$ time. 
\end{theorem}

The theorem immediately implies that LLDS+(3) has a factor-1.5 approximation as any feasible solution for $FT(3)$ would be a factor-1.5 approximation for LLDS+(3). In the following, we extend this trivial observation to have a factor-($1.5-O(\frac{1}{n})$) approximation for LLDS+(3).

\begin{corollary}
Let $S$ be a string of length $n$ such that each letter appears at most 3 times in $S$.
Then LLDS+(3) admits a polynomial-time approximation algorithm with a factor of $1.5-O(\frac{1}{n})$ if a feasible solution exists.
\end{corollary}

{\bf Proof.} 
First fix some constant (positive integer) $D~(D<|\Sigma|)$. Then for $t=1$ to $D$, we enumerate all the sets which contains letters appearing exactly 3 times in $S$. For a fixed $t$, let such a set be $F_t=\{a_1,a_2,...,a_t\}$. We put the 3-blocks $a_i^{(1)}a_i^{(2)}a_i^{(3)}$, $i=1..t$, in the solution. 
(If two such 3-blocks overlap, then we immediately stop to try a different set $F'_t$; and if all valid sets of size $t$ have been tried, we increment $t$ to $t+1$.) The substrings of $S$, between 
$a_i^{(1)}$ and $a_i^{(2)}$, and $a_i^{(2)}$ and $a_i^{(3)}$, will then be deleted.
Finally, for the remaining letters we use 2-SAT
to test whether all together, with the 3-blocks, they form a feasible solution (note that
$a_i^{(1)}a_i^{(2)}a_i^{(3)}$ will serve as an obstacle and no valid interval
for 2-SAT should contain it), this can be checked in $O(n^2)$ time following Theorem~2. Clearly, with this algorithm, either we compute the optimal solution with at most $D$ 3-blocks, or we obtain an approximate solution of value $2|\Sigma|+D$. Since {OPT} is at most $3|\Sigma|$, the approximation factor is
$$\frac{3|\Sigma|}{2|\Sigma|+D}=1.5-O(\frac{1}{|\Sigma|}),$$
which is $1.5-O(\frac{1}{n})$, because $|\Sigma|$ is at least $\lceil n/3\rceil$.
The running time of the algorithm is $O({|Sigma|\choose D}\cdot O(n^2))=O(n^{D+2})$, which is polynomial as long as $D$ is a constant.
\hfill $\Box$

In the next section, we show that if the LD-blocks are arbitrarily positively weighted, then the problem can be solved in $O(n^2)$ time. Note that the
$O(n)$ time algorithm in Section~2.1 assumes that the weight of any LD-block is its length, which has the property
that $\ell(s)=\ell(s_1)+\ell(s_2)$, where $s=s_1s_2$, $s_1$ and $s_2$ are LD-blocks on the same letter $x$, and $\ell(s)$ is the length of $s$ (or the total number of letters of $x$ in $s_1$ and $s_2$).

\section{A Dynamic Programming Algorithm for Weighted-LDS}

Given the input string $S=S[1...n]$, let $w_x(\ell)$ be the weight of LD-block $x^\ell, x\in\Sigma, 2\leq \ell\leq d$, where $d$ is the maximum number of times a letter appears in $S$. Here,
the weight can be thought of as a positive function of $x$ and $\ell$ and it does not even have to be increasing on $\ell$. For example, it could be that $w(aaa)=w_a(3)=8, w(aaaa)=w_a(4)=5$. Given $w_x(\ell)$ for all $x\in \Sigma$ and $\ell$, we aim to compute the maximum weight letter-duplicated string (Weighted-LDS) using dynamic programming.

Define $T(n)$ as the value of the optimal solution of $S[1...n]$ which contains the character $S[n]$. Define $w[i,j]$ as the maximum weight LD-block $S[j]^\ell$ ($\ell \geq 2$) starting at position $i$ and ending at position $j$; if such an LD-block does not exist, then $w[i,j]=0$. Notice that $S[j]^\ell$ does not necessarily have to contain $S[i]$ but it must contain $S[j]$.
We have the following recurrence relation.

\begin{align*}
T(0) & = 0, \\
T(i) & = \max\limits_{S[y] \neq S[i]}
\begin{cases}
T(y) + w[y+1, i] & \mbox{if } $w[y+1, i]$> 0,\\
0 & \text{otherwise.}
\end{cases}
\end{align*}

The final solution value is $\max\limits_{n} T(n)$. This algorithm clearly takes $O(n^2)$ time, assuming $w[i,j]$ is given. We compute the table $w[-,-]$ next.

    \noindent
    1. For each pair of $\ell$ (bounded by $d$, the maximum number of times a letter appears in $S$) and letter $x$, compute 
    $$w_x'(\ell) = \max
\begin{cases}
w_x'(\ell-1)&\\
w_x(\ell) &
\end{cases},$$ 
with $w_x'(1) = w_x(1)$.
This can be done in $O(d|\Sigma|)=O(n^2)$ time.
\newline

    \noindent
    2. Compute the number of occurrence of $S[j]$ in the range of $[i,j]$, $N[i,j]$. Notice that $i\leq j$ and for the base case we have $S[0] = \emptyset$.\\
    
    \begin{align*}
        N(0,0) & = 0, \\
        N(0,j) & = N(0,k) + 1, ~~k = \max
            \begin{cases}
            \{y|s[y]=s[j],1\leq y < j\} & ~\\
            0 &
            \end{cases}
    \end{align*}
    And,
   \begin{align*}
    N(i,j)  = 
        \begin{cases}
            N(i-1, j), & {if~ s[i-1]\neq s[j]}\\
            N(i-1,j)-1, & {if~ s[i-1] = [j]}
        \end{cases}
    \end{align*}
    
This step takes $O(n^2)$ time.
\newline
    
\noindent
3. Finally, we compute
    \begin{align*}
        w[i, j] = \begin{cases}
        w'_{s[j]} (N(i,j)), & if~N(i,j) \geq 2 \\
        0, & else
    \end{cases}
    \end{align*}
    This step also takes $O(n^2)$ time.
We thus have the following theorem.
\begin{theorem}
Let $S$ be a string of length $n$ over an alphabet $\Sigma$ and $d$ be the maximum number of times a letter appears in $S$. Given the weight function $w_x(\ell)$ for $x\in\Sigma$ and $\ell\leq d$, the maximum weight letter-duplicated subsequence (Weighted-LDS) of $S$ can be computed in $O(n^2)$ time. 
\end{theorem}

We can run a simple example as follows. Let $S=ababbaca$. Suppose the table
$w_x(\ell)$ is given as Table~1.
\begin{table}
	\centering
\caption{Input table for $w_x(\ell)$, with $S=ababbaca$ and $d=4$.} 
	\begin{tabular}{|c|c|c|c|c|}
\hline $x\backslash\ell$ & 1 & 2 & 3 & 4 \\
\hline $a$ & 5 & 10 & 20 & 15 \\
\hline $b$ & 4 & 16 & 8 & 3\\
\hline $c$ & 1 & 3 & 5 & 7 \\
\hline
	\end{tabular}
\end{table}
At the first step, $w_x'(\ell)$ is the maximum weight of a LD-block made with $x$ and of length at most $\ell$. The corresponding table $w_x'(\ell)$ can be computed as Table~2.
\begin{table}
	\centering
\caption{Table $w_x'(\ell)$, with $S=ababbaca$ and $d=4$.} 
	\begin{tabular}{|c|c|c|c|c|}
\hline $x\backslash\ell$ & 1 & 2 & 3 & 4 \\
\hline $a$ & 5 & 10 & 20 & 20 \\
\hline $b$ & 4 & 16 & 16 & 16\\
\hline $c$ & 1 & 3 & 5 & 7 \\
\hline
	\end{tabular}
\end{table}
At the end of the second step, we have Table~3 computed. 
\begin{table}
	\centering
\caption{Part of the table $N[i,j]$, with $S=ababbaca$ and $d=4$.} 
	\begin{tabular}{|c|c|c|c|c|c|c|c|c|}
\hline $i\backslash j$ & 1 & 2 & 3 & 4 & 5 & 6 & 7 & 8 \\
\hline $8$ & 0 & 0 & 0 & 0 & 0 & 0 & 0 & 1\\
\hline $\cdots$ & $\cdots$ & $\cdots$ & $\cdots$ & $\cdots$ & $\cdots$ & $\cdots$ & $\cdots$ & $\cdots$\\
\hline $3$ & 0 & 0 & 1 & 1 & 2 & 2 & 1 & 3\\
\hline $2$ & 0 & 1 & 1 & 2 & 3 & 2 & 1 & 3 \\
\hline $1$ & 1 & 1 & 2 & 2 & 3 & 3 & 1 & 4 \\
\hline
	\end{tabular}
\end{table}
From Table~3, the table $w[-,-]$ can be easily computed and we omit the details. For instance, $w[1,-]=[0,0,10,16,16,20,0,20]$. With that, the optimal solution value can be computed as $T(8)=36$, which corresponds to the optimal solution $aabbaa$.

\section{Concluding Remarks}
We consider the constrained longest letter-duplicated subsequence (LLDS+) and the corresponding feasibility testing (FT) problems in this paper, where all letters in the alphabet must occur in the solutions. We parameterize the problems with $d$, which is the maximum number of times a letter appears in the input sequence. For convenience, we summarize the results one more time in the following table. Obviously, we have many open problems.

\begin{table}
	\centering
\caption{Summary of results on LLDS+ and FT, the ? indicates that the problem is still open.} 
	\begin{tabular}{|c|c|c|c|}
\hline $d$ & \emph{LLDS+}$(d)$ & $FT(d)$&Approximability~of~\emph{LLDS+}$(d)$\\
\hline $d\geq 6$ & NP-hard & NP-complete & No~approximation \\
\hline $d=3$ & ? & P & 1.5-$O(\frac{1}{n})$\\
\hline $d=4,5$ & ? & ? & ? \\
\hline
	\end{tabular}
\end{table}

We also consider the weighted version (without the `full-appearance' constraint), for which we give a non-trivial $O(n^2)$ time dynamic programming solution. 

If we stick with the `full-appearance' constraint,
one direction is to consider two additional variants of the problem where the solutions must be a subsequence of $S$, in
the form of $x_1^{d_1}x_2^{d_2}\cdots x_k^{d_k}$ with $x_i$ being a substring (resp. subsequence) of $S$ with length at least 2, $x_j\neq x_{j+1}$ and $d_i\geq 2$ for all $i$ in $[k]$ and $j$ in $[k-1]$. 
Intuitively, for many cases these variants could better capture the duplicated patterns in $S$.
At this point, the NP-completeness results (similar to Theorem 1 and Corollary 1) would still hold with minor modifications to the proofs. (This reduction is still from $(\leq 2,1,\leq 3)$-SAT and is additionally based on the following fact: given a {\em (2,1)-sequence} $T=ABCCAB$ over $\{A,B,C\}$, where $A,B$ and $C$ all appear twice, the corresponding maximal `substring-duplicated-subsequences' or `subsequence-duplicated-subsequences' of $T$ are $ABAB=(AB)^2$ or $CC$.) But whether these extensions allow us to design good approximation algorithms needs further study.
Note that, without the `full-appearance' constraint, when $x_i$ is a subsequence of $S$, the problem is a generalization of Kosowski's longest square subsequence problem \cite{DBLP:conf/spire/Kosowski04} and can certainly be solved in polynomial time.


\bibliographystyle{plainurl}
\bibliography{main}

\end{document}